# Multi-Connected Ontologies

Philip Davies
Higher Education
Bournemouth and Poole College
Bournemouth, UK
pdavies@bpc.ac.uk

David Newell
Software Systems Research Group
Bournemouth University
Bournemouth, UK
dnewell@bournemouth.ac.uk

Abigail Davies
St Johns College
Oxford University
Oxford, UK
abigail.davies@sjc.ox.ac.uk

Damla Karagözlü
Software Systems Research Group
Bournemouth University
Bournemouth, UK
<damla.karagozlu@gmail.com>

*Abstract* – **Ontologies have been used for the purpose of bringing system and consistency to subject and knowledge areas. We present a criticism of the present mathematical structure of ontologies and indicate that they are not sufficient in their present form to represent the many different valid expressions of a subject knowledge domain. We propose an alternative structure for ontologies based on a richer multi connected complex network which contains the present ontology structure as a projection. We demonstrate how this new multi connected ontology should be represented as an asymmetric probability matrix.**

I. INTRODUCTION

**The present state of ontologies**
There has been exceptional growth in the annotation of information prompted by the increasing need to share data and study objects based on their structure and semantics. (Gruber 1993) We now find annotated information in a wide range of areas such as language, biology, computing, medicine, web content, etc. Annotated information is created from structured vocabularies known as ontologies. Many disciplines have now developed their own standardized ontologies to enable the sharing of information in their fields. SNOMED, for instance has been produced in the field of medicine, (Price and Spackman 2000) as well as many others which are now being referenced (Noy and McGuinness n.d.).

An ontology defines a common vocabulary for researchers who need to share information in a domain. Many subject areas are now developing ontologies so that specialists can share information in their fields not only with other specialists but even with machines. (Protégé n.d.) Machine-interpretable definitions of basic concepts in the domain and relations among them enable the widespread use of information on the internet and the construction of expert systems.

An ontology uses relationships to organize concepts into hierarchies or subject domains. (Noy and McGuinness n.d.) This paper investigates the present structure of ontologies and whether they are applicable to describing subject domains in their present form. The basic problem we consider is whether the present structure of ontologies is rich enough to represent subject domains fully. We contend that the concept of ontologies needs to be extended in order to fully realise a complete subject domain and we indicate ways in which this extension might be approached

**Critique of Ontologies**
Our approach to ontology structure is drawn from the ideas of the German philosopher Martin Heidegger (1889 – 1976). Heidegger was critical of a one-dimensional division of the world into simplistic categories. According to Heidegger, "*The philosophical tradition has misunderstood human experience by imposing a subject-object schema upon it.*" (Blatner 2006)

Heidegger gives the example of a hammer which cannot be represented just by its physical features and functions. To understand the hammer you cannot detach it from its relationship to the nails, to the anvil, to the wood, to the experience and skill of the carpenter or to a hundred other things. Just putting it in a category of tools, in an ontology cannot fully capture the human idea of the object and its role in the world. A more complex structure is need to capture the representation of reality. (Blatner 2006)

Robert Pirsig (Pirsig 1974) has also made the point that there always appears more than one workable hypothesis to explain a given phenomenon, and that the number of possible hypotheses appears unlimited. He has developed the idea that there are two types of thinking, the classical and the romantic. The classical way of thinking is characterised by analysing things into their component parts, whereas the romantic sees things as a whole. Classical thought would analyse an object like a motorbike into its physical components; nuts bolts etc. but you can also analyse the motor bike into its functional parts: heat exchanger, generator, exhaust system etc. Pirsig points out that each analysis is equally valid but produces different results. It depends on how you wield the knife of analysis to separate part from part. For example if you take a cylindrical chunk of clay you can cut it straight down and the product is circles, but if you decide to cut at an angle the result is ellipses, if you cut horizontally you obtain rectangles. The result of any analysis is also the product of what you decide to do and how you decide to cut, as much as it is a product of the artefact you are looking at. No analysis is unique.

This directly affects the construction of ontologies as these are the results of detailed analysis of a subject area. Since different analyses lead to different ontologies and each may be equally valid, it has become necessary to agree on a convention as to what the structure of any given ontology may be and this agreement by subject experts is the method chosen to determine an agreed ontology. But we contend here that agreement by convention on the structure of a subject ontology is not sufficient, as there are intrinsic differences between representations which cannot be reconciled because the subject domain is richer than any single ontology can capture. Different ontologies result from on the way the knife has been wielded as much as the subject area itself.

David Bohm in his book Wholeness and the Implicate Order (Bohm 1980) presents a critique of the fragmentation that classical thought has introduced into our description of the world. He says that it has always been necessary and appropriate to divide things up and separate them in order to reduce problems to manageable proportions but in so doing we lose sight of the whole. In dividing things up we make the mistake of thinking that the fragments we produce are a proper description of the world as it is. The problem is that there are many different ways of thinking about something and of categorising concepts and ideas. And no one way is better than another. He uses the field of quantum mechanics to illustrate this with its wave picture and particle picture of reality which are at the same time incompatible and indivisible. *"All our different ways of thinking are to be considered as different ways of looking at the one reality"* says Bohm. (Bohm 1980) Each view gives only one appearance of the object in some respect. *"The whole object is not perceived in any one view but rather it is grasped only implicitly as that single reality which is shown in all these views."* (Bohm 1980)

This has direct application to the way we use ontologies. These are constructed on the premise that in order to communicate about a particular subject domain unambiguously we need to have an agreed reference point, the ontology, which fixes precisely and unambiguously the component of the subject domain and its fixed relationships to other points. What Heidegger, Pirsig and Bohm are telling us is that this approach may be wrong from the outset and ultimately unachievable in the long term. A single ontology to describe the whole of reality is not something that exists. Rather many incompatible ontologies will exist that are equally valid descriptions of reality. And merely agreeing on one of them for the sake of convention will not enable a full picture of the reality to be represented. What is needed is a larger concept which contains all possible ontologies in a single undivided structure implicitly and from which they can be explicated.

We can liken the new structure to a three dimensional object that casts different shadows depending on which way the light falls and each shadow represent the ontology while the object is the reality.

## II. A New Approach to Constructing Ontologies

We propose to adopt a new approach to describing knowledge systems based on the idea that there is no one correct method of organising a subject domain in an ontology but rather there are many different ontology structures that adequately and correctly represent a body of knowledge. Each ontology gives only one appearance of the subject in some particular respect.

We will consequently seek to develop an approach to multi-ontologies and suggest a way in which they can be connected together in to a larger multi connected ontology. We will consider four stages in the construction of any knowledge system.

Stage 1 Introducing Order
Stage 2 Introducing Coherence
Stage 3 Introducing Proximity
Stage 4 Introducing Co-Requisites and Pre-Requisites

These four stages will lead to a larger concept for ontologies that encompass the present understanding of ontologies as a subset.

## III. Stage 1 Introducing Order

Ontologies specify the structure and relationships within a body of knowledge. Usually ontologies are represented as knowledge hierarchies with the most general concepts at the top and more detailed and specific concepts at lower levels (Bhattacharya 6 Mar 2010). Thus a body of knowledge is divided into sections, sub-sections, sub-sub-sections etc. The structure of these knowledge hierarchies is naturally representable as networks, where each node on the network represents a unit of knowledge and where the relationship of each part to every other is determined and specified within the ontology. An ontology can be represented as a tree network where there is a maximum of one path between any two nodes.

We may adopt an addressing system which corresponds to this knowledge hierarchy where each address is correspondingly specified by sections, sub-sections, sub-sub-sections etc. See Figure 2

The advantage of the simple tree model is that the number of hops from the root provides the level of the node. The disadvantage is that the structure does not contain the ordering of the concepts.

However because of the hierarchical nature of sections, subsection etc, the ontology has an implicit ordering. The structure of an ontology is built up from fragments of knowledge which have an order determined by their pre-requisites. Consequently the simple tree depicted in Figure 2 is not sufficient to model this structure as it lacks the necessary order. We use an ordered tree for this description where the branches from each node are ordered so that the sub-nodes have an order of preference. (Newman 2010)

**Principle 1: Simple ontologies are to be represented mathematically as an ordered tree**

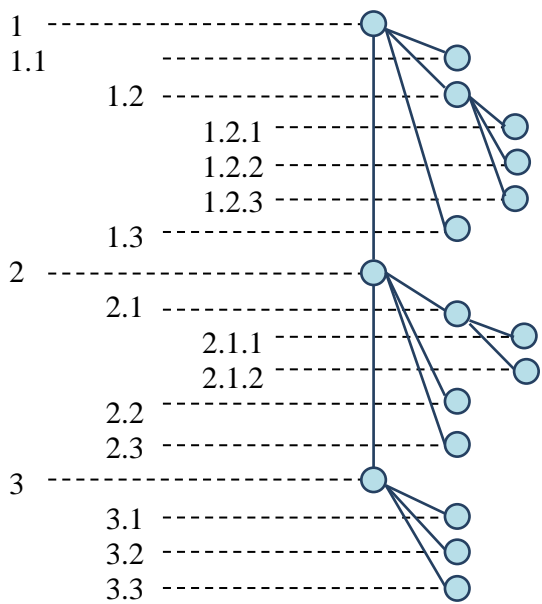

**Figure 2: Knowledge hierarchy corresponding to an unordered tree**

The ordered tree network is distinguished by
1. there is a maximum of one route from any node to any other node
2. Branches from any given node have an implicit order.

These two properties ensure that the ordered tree network has the necessary properties to represent simple knowledge categorisation and sub-categorisation within an ontology. This structure will also enable a wide variety of knowledge maps to be represented. (Davies 2011)

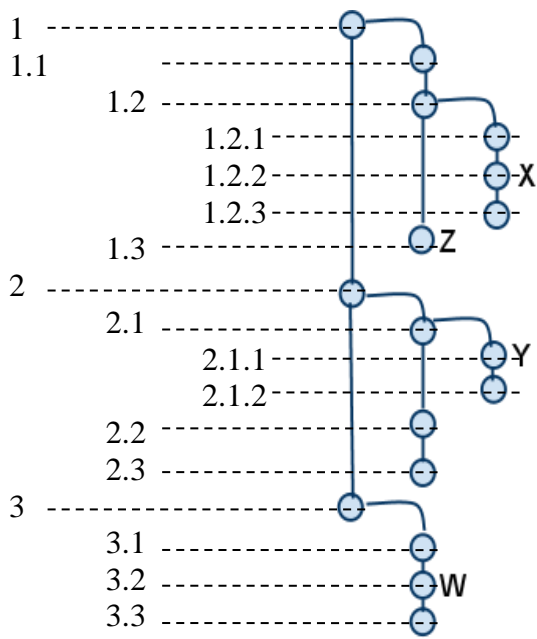

**Figure 4: Knowledge hierarchy corresponding to an ordered tree**

IV. STAGE 2 INTRODUCING COHERENCE

We start with the recognition that no one ontology is the correct or the ultimate expression of a subject domain and accept that there are many different ontologies which all adequately represent the knowledge area from different points of view. This is a significant departure from the present understanding of ontologies and we therefore present it as our next principle.

**Principle 2:** The same structure can be analysed in different ways if it is complex enough

We next recognise that all these different ontologies are ordered trees which mathematically can be combined into a more complex network containing each of them as a sub network. We therefore introduce a multi connected ontology represented by a mesh network containing multi-connected pathways between nodes. This extends the model for the ontology from that of a tree to a mesh network.

Using Bohm's terminology we would say that the multi connected ontology is the implicate order while a particular decomposition ontology is the explicate order. (Bohm 1980)

That means that starting from a larger mesh network we can generate an ordered tree by breaking certain connections in the complex structure effectively decomposing it into a simpler ordered tree. In this way the implicate order of the multi connected ontology becomes the explicate order of the simple ontology or the ordered tree. The breaking of different links in the multi connected ontology will produce a different ontology.

**Principle 3:** A multi connected ontology can be decomposed into at least one simple ordered tree

**Principle 4:** Different decompositions produce different but equally valid ontologies

In this way you can unloose or break certain connections in a full multi-connected network which will lead to a one decomposition that produces a certain ontology, while another method of breaking connections will lead to another decomposition and a different ontology of the same reality.

Links can be variable because different items of knowledge can be linked together in different ways. What doesn't vary is the items of knowledge themselves. The content of the knowledge must remain invariant but one item can precede another or follow another depending on presentation and other factors. At a lower level each knowledge item or ontology node may be explained in many different ways.

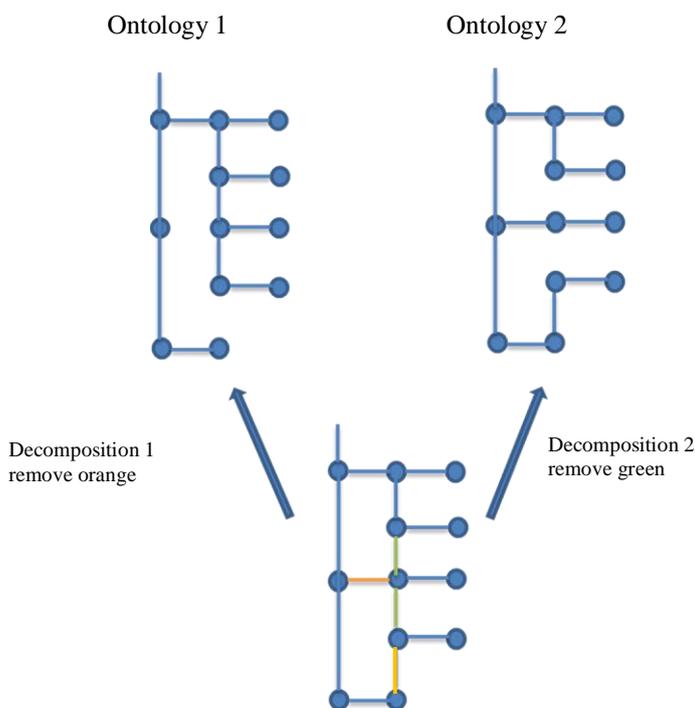

**Figure 5: Decomposing the multi connected ontology**

For instance binary arithmetic can be introduced in a variety of ways, but whichever way is chosen it is still teaching the same thing. That is because the content has not varied and the content is determined by the nodes. What is determined by the links is the presentation. Links within ontologies represent the way of explaining the knowledge or packaging the knowledge for student consumption or opening up the subject. All this information is contained in the links. One tutor may adopt a different approach to another by which we mean he will present the nodes in a different order. So each node has a different number of presentations but the content is the same. This is the basic difference between knowledge and education. Knowledge of a subject is the acquisition of a node but the node can be delivered in many different ways and the delivery is concerned with education.

The same relation exists between teaching and learning. Learning is fixed on the acquisition of knowledge nodes, while teaching is involved in the arrangement of the knowledge nodes in a form the tutor presents them. Each tutor may be different and present the knowledge in a different way – yet they are all teaching the same knowledge.

Each presentation may be different and based on different learning styles or learning needs. There may be different degrees of information required where the weak student needs a lot of information and the strong student needs very little. This will define the difference between weak and strong in the student model.

**Decompositions**
We can formally express decompositions using the adjacency matrix. Let $M_{ij}$ be the adjacency matrix of the multi-connected ontology and let $O_{ij}$ be the particular decomposition tree ontology. Then

$$X_{ij} M_{ij} = O_{ij}$$

where $X_{ij}$ is the decomposition operator. In effect $X_{ij}$ takes the multi-connected $M_{ij}$ into a specific tree $O_{ij}$ which represents the structure and organisation of the knowledge as presented by a particular tutor for a particular student at a particular time with a particular level of subject knowledge. $X_{ij}$ is thus a function of all these parameters.

$X_{ij}$ exists only if $M_{ij}^{-1}$ exists since:

$$X_{ij} M_{ij} M_{ij}^{-1} = O_{ij} M_{ij}^{-1}$$
$$X_{ij} = O_{ij} M_{ij}^{-1}$$

Maximally connected networks (where all nodes are connected to all other nodes) have a simple adjacency matrix $K_{ij}$ in which every component is equal to 1 except for the diagonal components which are equal to 0.

$K_{ij} = 1$ (for $i \neq j$) and
$K_{ij} = 0$ (for $i = j$)

All $K_{ij}$ of dimension n have an inverse $K_{ij}^{-1}$ which is given by

$K_{ij}^{-1} = 1/(n-1)$ (for $i \neq j$) and
$K_{ij}^{-1} = -(n-2)/(n-1)$ (for $i = j$)

The existence of the inverse means that every decomposable ontology can be generated from the maximally connected network.

The inverse of $X_{ij}$ will be $X_{ij}^{-1}$ which will restore the global multi-connected network from the specific tree

$$X_{ij}^{-1} X_{ij} M_{ij} = X_{ij}^{-1} O_{ij}$$
$$M_{ij} = X_{ij}^{-1} O_{ij}$$

The existence of the inverse is important because it means that given a particular knowledge decomposition we can always get to any other knowledge decomposition via the multi-connected ontology.

This may be clearer if we take a particular decomposition as an example. Consider the ten node network shown in Figure 6

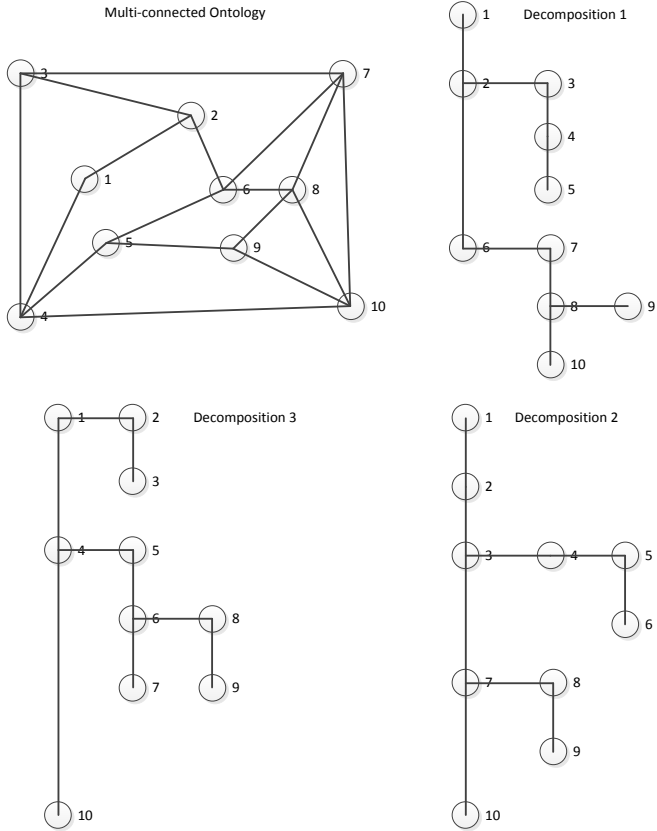

**Figure 6: Ten node network decomposition example**

The multi connected ontology can be decomposed in a number of ways, three of which are illustrated. For clarification we have numbered the ten nodes consecutively from 1 to 10 and the Adjacency matrix $M_{ij}$ is shown in Figure 7

**Figure 7: Adjacency matrix Mij**

The adjacency matrix $M_{ij}$ has an inverse $M_{ij}^{-1}$ which takes the form shown in Figure 8

| 1 | 3/2 | -1/2 | -1/2 | -1 | -1/2 | -1 | 1/2 | 1 | 1/2 |
|---|---|---|---|---|---|---|---|---|---|
| 3/2 | 0 | -1 | 0 | -1/2 | 1/2 | 0 | 1/2 | -1/2 | 0 |
| -1/2 | -1 | 0 | 1 | 1/2 | 1/2 | 1 | -1/2 | -3/2 | 0 |
| -1/2 | 0 | 1 | 0 | 1/2 | -1/2 | 0 | -1/2 | 1/2 | 0 |
| -1 | -1/2 | 1/2 | 1/2 | 1 | 1/2 | 0 | -1/2 | 0 | -1/2 |
| -1/2 | 1/2 | 1/2 | -1/2 | 1/2 | 0 | 0 | 0 | 1/2 | 0 |
| -1 | 0 | 1 | 0 | 0 | 0 | 0 | 0 | 0 | 0 |
| 1/2 | 1/2 | -1/2 | -1/2 | -1/2 | 0 | 0 | 0 | 1/2 | 1/2 |
| 1 | 1/2 | -3/2 | 1/2 | 0 | 1/2 | 0 | 1/2 | -1 | 1/2 |
| 1/2 | 0 | 0 | 0 | -1/2 | -1/2 | 0 | 1/2 | 1/2 | 0 |

**Figure 8: Inverse adjacency matrix $M_{ij}^{-1}$**

The adjacency matrix of decomposition 1 $O_{ij}(1)$ which is a particular instance of an ontology of the multi-connected ontology $M_{ij}$ is given by Figure 9 where the greyed out boxes indicate those components of $M_{ij}$ which are to be removed.

**Figure 9: The adjacency matrix of decomposition 1 $O_{ij}(1)$**

The adjacency matrix of decomposition 2 $O_{ij}(2)$ which is another particular instance of an ontology of the multi-connected $M_{ij}$ is given by Figure 10.

**Figure 10: The adjacency matrix of decomposition 2 $O_{ij}(2)$**

The adjacency matrix of decomposition 3 $O_{ij}(3)$ which is another particular instance of an ontology of the multi-connected ontology $M_{ij}$ is given by Figure 11.

**Figure 11: The adjacency matrix of decomposition 3 $O_{ij}(3)$**

It follows from the preceding that not only can multi connected ontologies be decomposed into 'instance ontologies' as we may call a standard ontology but conversely a multi connected ontology can be constructed from instance ontologies and they can be combined into a mesh network. The converse of Principle 3 follows.

**Principle 5:** A multi connected ontology can be constructed from simple instance ontologies

Thus the two ontologies in Figure 3 can be constructed from the larger mesh network by the selection of the correct links. However there must be the same nodes for this to work.

**Principle 6** for two ontologies to be identical they must have identical nodes, though not necessary identical links

It is quite easy to prove that any two trees of equal number of nodes but different links could be combined into a single multi-connected ontology. Consider the adjacency matrix of the two complementary trees, call them A and B, then it is always possible to form a new adjacency matrix C such that

$$C = A \oplus B$$

Where we have defined ⊕ as the operator which adds two matrix elements together according to the rule:

$M_{ij} \oplus N_{ij} = 1$ (if $M_{ij}$ and/or $N_{ij} = 1$)
$M_{ij} \oplus N_{ij} = 0$ (if $M_{ij}$ and $N_{ij} = 0$)

This C will be representable as a new network, which is not a tree.

Indeed we can go further and state that a full maximal multi-connected system of n nodes $K_n$ where every node is connected to every other node can be decomposed into any tree structure of n nodes $T_n$ and that all $T_n$ are subsets of $K_n$

$T_n \subseteq K_n$

In general every ontology could be decomposed from the maximal multi connected network.

However we need to be aware that some systems do not yield to this simple analysis as they are not based on different links but on different nodes.

Principle 6 is the fundamental principle that puts a difference between what we are doing and what is being done elsewhere as the structure of an ontology is usually rigidly defined not only by its nodes but also by the fixed links that relate those nodes, so that if the links change then the ontology changes. In Principle 6, we are saying that this is not necessarily so, or that the two ontologies are equivalent, even though they may have different structures. However, the number of nodes must be the same in all cases as they represent knowledge elements and ontologies with different knowledge elements contain different knowledge areas. This is worth restating again.

> Two ontologies are the same if they have the same nodes but not necessarily the same links

## V.  STAGE 3 INTRODUCING PROXIMITY

There are many ways to arrange the nodes of a subject ontology. For instance, if we take the example of computing as a subject area, the knowledge nodes can be arranged in thematic order, logical order, functional order, historical order, geographical order etc. There is no end to the number of ways that knowledge nodes can be linked and presented, other than the mathematical limit of the total number of ways of arranging a finite number of nodes which is n!/2 since the number of ways of arranging n distinguishable objects is n! and we are treating reverse orders as the same arrangement for tree networks.

One way of doing this is to make each of these decompositions dependent on a set of decomposition parameters which determines the ordering. To do this each subject node would need to be tagged with these meta-subject parameters so that each node carries with it the information about its order in history or geography or function etc. However this is not needed if we use the decomposition operator $X_{ij}$ as all the information as to the structure will reside here. Thus there will be decomposition operators which will represent the different structures. We could speak of a Historical decomposition $X_{ij}(H)$ or a geographical decomposition $X_{ij}(G)$ etc. We can generalise this to $X_{ij}(k)$ In reality there will be a maximum of n!/2 such possible decompositions for a subject area with n nodes.

These decompositions are individually constructed (as are ontologies themselves) by individual subject experts who may be expected to provide their own decompositions very much as different experts would produce different books with different contents structures even though they were writing on the same subject as another expert. Each expert arranges his material in his own way and in a way that suits him and his way of thinking and presenting information. We may speak therefore of individual tutor or expert decompositions $X_{ij}(E_k)$ corresponding to their understanding of how the subject information should be arranged and presented. Hence

$X_{ij}(E_k) M_{ij} = O_{ij}(E_k)$

Where $O_{ij}(E_k)$ is the ontology produced by Expert $E_k$ A full determination of $E_k$ will require a tutor model with a full set of identified parameters. Similarly there will be a preferred decomposition for a particular student S who will have his own level of pre-existing knowledge, speed of acquisition of new knowledge etc. The full determination of this will require a student model with a full set of defined parameters. The full details of the tutor model, student model and other models will be dealt with in a separate paper.

The consequence of moving from a tree to a mesh network is that we now have more than one route between any two nodes. Hence within the multi connected ontology $M_{ij}$ there are multiple routes between nodes and not all paths will be equal. Some paths will be very common and chosen by a majority of experts. Some paths may be much rarer and chose by perhaps only one expert. The accumulated frequency of choice may be interpreted as a probability value which indicates the likelihood of one node being linked to another by the creators of the separate ontologies for each decomposable ontology created by an individual expert or tutor.

Consequently some subject nodes will have a higher probably of transition within the domain than other subject nodes and can be thought of as being 'closer' to each other for that reason. If there is more than one route away from a subject node then each pathway will be weighted according to the probability that an expert may move from one to another. We will model this by introducing probabilities into the adjacency matrix by replacing the 1s with probability values between 0 and

1 where 0 indicates no probability of a transition between two nodes and 1 indicates a 100% probability which means that one node must lead to another.

In this way the adjacency matrix from Figure 7 would be transformed to something like Figure 12

|   | .2 |   | .8 |   |   |   |   |   |   |
|---|---|---|---|---|---|---|---|---|---|
| .2 |   | .2 |   |   | .6 |   |   |   |   |
|   | .2 |   | .1 |   |   | 1 |   |   |   |
| .8 |   | .1 |   | .05 |   |   |   |   | .05 |
|   |   |   | .05 |   | .2 |   |   | .2 |   |
|   | .6 |   |   | .2 |   | 1 | .4 |   |   |
|   |   | 1 |   |   | 1 |   | .4 |   | .1 |
|   |   |   |   |   | .4 | .4 |   | .2 | 1 |
|   |   |   |   | .2 |   |   | .2 |   | .5 |
|   |   |   | .05 |   |   | .1 | 1 | .5 |   |

**Figure 12: Probability Adjacency Matrix**

The problem of finding a suitable pathway through the multi connected ontology which maximises the probability of transition then reduces to a travelling salesman type problem.

### VI. STAGE 4 INTRODUCING CO-REQUISITES AND PRE-REQUISITES

Pre-requisite knowledge domains indicate that one area of subject knowledge must be taught prior to another. This is a consequence of knowledge building on previous knowledge. Therefore within our model a mechanism is required to show which subject knowledge nodes are prior to other nodes. Pre-requisites mean that one subject node must come before another.

The concept of pre-requisites introduces the notion of direction into the ontology network. Directed networks only allow one route between two knowledge nodes and are usually represented by arrows. To represent this in our adjacency matrix we will introduce asymmetry into the adjacency matrix to show that connections are just one way.

In this way the adjacency matrix from Figure 12Figure 7 would be transformed to something like Figure 13

|   | .2 |   | .8 |   |   |   |   |   |   |
|---|---|---|---|---|---|---|---|---|---|
|   |   | .2 |   |   | .6 |   |   |   |   |
|   | 0 |   | 0 |   |   | 1 |   |   |   |
| .8 |   | .1 |   | 0 |   |   |   |   | .05 |
|   |   |   | .05 |   | .2 |   |   | .2 |   |
|   | .6 |   |   | .2 |   | 1 | .4 |   |   |
|   |   | 1 |   |   | 1 |   | .4 |   | .1 |
|   |   |   |   |   | 0 | .4 |   | .2 | 1 |
|   |   |   |   | .2 |   |   | .2 |   | .5 |
|   |   |   | .05 |   |   | 0 | 0 | 0 |   |

**Figure 13: Directed Probability Adjacency Matrix**

**Student Adaption Models**
The idea of directed networks allows us to more closely control the direction of learning that a student may have to undertake in studying a particular knowledge domain. Not all students will take the same path. For some students some nodes will not be necessary for their personalised learning, for other students they will be. (Boyle 2003) Thus the directed pathways, representing prerequisites of learning will be different for different students. This means that the adaption necessary of each individual student is not based on the nodes that they need to acquire. The nodes give you the level of knowledge (GCSE, A level, degree, masters) the more nodes the more knowledge acquired. Adaption on the other hand is based on the presentation of knowledge. (N. a. Rowe 2011) (N. C. Rowe 2010 ) Knowledge presentations are tailored for each individual student according to their ability or their pre-knowledge or how rapidly they can take on board knowledge (comprehension) or whether they are weak or strong students or their learning styles or their completeness, speed angle (theoretical/hands on) (Chen 2007).

**Tutor Presentation Models**
Presentations can complement each other. A student may learn the theory first and then apply the learning or consolidate it with a practical task or vice versa. Presentations take the knowledge and wrap it up in a particular way and make it accessible to the student. Presentations are independent of the knowledge they package. They are the vehicle for taking the knowledge to the student. The presentation has to decompose the knowledge into its components and then present them in a logical form that builds on what has gone before. They should be presented in a way that suits the student. Consequently you have to start with where the student is and build a bridge from there. Thus the student model needs to take account of the initial position of the student and encode that in the student signature. (Cutts 2009) and (Newell 2011).

### VII. CONCLUSION

We have extended the concept of ontology to include multiple representations of a knowledge domain.

The full representation of a model of a knowledge domain made from subject nodes connected into a particular structure requires an ordered multi-connected network described by an asymmetric probability adjacency matrix

In future, we need to construct a full, robust tutor model and student model to automate the learning process. Our vision is to build this into a novel abstract conceptual data model encompassing all the properties that are needed to make explicit the qualities of an effective learning method of which the structure of ontologies is but one part. (McGreal 2004)

Finally, although work discussed in this paper answered research questions posed in previous papers, it has indicated further questions with a different emphasis: